\documentclass[prb,showpacs,preprintnumbers,preprint,amsmath,amssymb]{revtex4}
\usepackage{amssymb}
\usepackage{graphicx}
\usepackage{dcolumn}% Align table columns on decimal point
\usepackage{bm}% bold math
\newcommand{\SmOsSb}{SmOs$_{4}$Sb$_{12}$}
\newcommand{\LaOsSb}{LaOs$_{4}$Sb$_{12}$}
\newcommand{\YOsSb}{YOs$_{4}$Sb$_{12}$}
\newcommand{\LuOsSb}{LuOs$_{4}$Sb$_{12}$}
\newcommand{\PrFeSb}{PrFe$_{4}$Sb$_{12}$}
\newcommand{\PrOsSb}{PrOs$_{4}$Sb$_{12}$}
\newcommand{\YbFeSb}{YbFe$_{4}$Sb$_{12}$}
\newcommand{\LaFeP}{LaFe$_{4}$P$_{12}$}
\newcommand{\SmFeP}{SmFe$_{4}$P$_{12}$}
\newcommand{\SmRuP}{SmRu$_{4}$P$_{12}$}
\newcommand{\SmOsP}{SmOs$_{4}$P$_{12}$}

\newcommand{\SmFeSb}{SmFe$_{4}$Sb$_{12}$}
\newcommand{\OsSb}{OsSb$_{2}$}
\newcommand{\CeOsSb}{CeOs$_{4}$Sb$_{12}$}
\newsymbol\grtsim 1326

\begin{document}

\title{Heavy Fermion Behavior, Crystalline Electric Field Effects, and Weak Ferromagnetism in \SmOsSb}

\author{W. M. Yuhasz}
\author{N. A. Frederick}
\author{P.-C. Ho}
\author{N. P. Butch}
\author{\\B. J. Taylor}
\author{T. A. Sayles}
\author{M. B. Maple}

\affiliation{Department of Physics and Institute for Pure and
Applied Physical Sciences, University of California San Diego, La
Jolla, CA 92093}

\author{J. B. Betts}
\author{A. H. Lacerda}
\affiliation{National High Magnetic Field Laboratory/Los Alamos
National Laboratory, \\Los Alamos, NM 87545}

\author{P. Rogl}
\affiliation{Institut f\"ur Physikalische Chemie, Universit\"at
Wien, A-1090 Wien, W\"ahringerstr. 42, Austria}

\author{G. Giester}
\affiliation{Institut f\"ur Mineralogie und Kristallographie,
Universit\"at Wien, A-1090 Wien, Althanstr. 14, Austria}
\date{\today}

\begin{abstract}

The filled skutterudite compound \SmOsSb{} was prepared in single
crystal form and characterized using x-ray diffraction, specific
heat, electrical resistivity, and magnetization measurements. The
\SmOsSb{} crystals have the \LaFeP-type structure with lattice
parameter $a = 9.3085$ \AA. Specific heat measurements indicate a
large electronic specific heat coefficient of \mbox{$\approx 880$
mJ/mol K$^{2}$}, from which an enhanced effective mass $m^{*}
\approx 170\ m_\mathrm{e}$ is estimated. The specific heat data
also suggest crystalline electric field (CEF) splitting of the
Sm$^{3+} J = 5/2$ multiplet into a $\Gamma_{7}$ doublet ground
state and a $\Gamma_{8}$ quartet excited state separated by $\sim
37$ K. Electrical resistivity $\rho$(T) measurements reveal a
decrease in $\rho$(T) below $\sim 50$ K that is consistent with
CEF splitting of \mbox{$\sim 33$ K} between a $\Gamma_{7}$ doublet
ground state and $\Gamma_{8}$ quartet excited state.  Specific
heat and magnetic susceptibility measurements display a possible
weak ferromagnetic transition at $\sim 2.6$ K, which could be an
intrinsic property of \SmOsSb{} or possibly due to an unknown
impurity phase.

\end{abstract}


\maketitle

\section{Introduction}

The family of filled skutterudite compounds exhibits a variety of
interesting strongly correlated electron phenomena and have
potential for thermoelectric applications. These phenomena
include, for example, superconductivity in
\LaFeP{},\cite{Meisner81} heavy fermion superconductivity in
\PrOsSb{},\cite{Bauer02a} ferromagnetism in
\SmFeP{},\cite{Takeda03a} Kondo insulating behavior in
\CeOsSb{},\cite{Bauer01} and valence fluctuations in
\YbFeSb{}.\cite{Dilley98} The filled skutterudites have the
chemical formula MT$_4$X$_1$$_2$ where M = alkali metal,
alkaline-earth, lanthanide, or actinide (Th or U); T = Fe, Ru, or
Os; X = P, As, or Sb. The unit cell for these compounds consists
of $34$ atoms crystallized in a \LaFeP-type structure with the
Im\={3} space group.\cite{Jeitschko77}

The only Sm-based filled skutterudites to be characterized at low
temperatures are all three phosphides and \SmFeSb{}; of these,
only \SmFeP{} has been studied in single crystal
form.\cite{Takeda03a,Jeitschko00,Sekine00,Takeda03b,Giri03,Fujiwara03,
Yoshizawa03,Matsuhira02,Danebrock96}  These studies have revealed
that \SmFeP{} is ferromagnetic below $1.6$ K, \SmRuP{} has a
metal-insulator transition at $16$ K, \SmOsP{} is an
antiferromagnet below $4.6$ K, and \SmFeSb{} is ferromagnetic
below $45$ K.  Experiments on \SmFeP{} have also uncovered heavy
fermion and Kondo-lattice behavior, with a Kondo temperature of
about $30$ K and an electronic specific heat coefficient of $\sim
370$ mJ/mol K$^{2}$.\cite{Takeda03a,Takeda03b}  By $7$ T, the
magnetization of \SmFeP{} at $1.8$ K does not saturate and only
reaches a value of $0.15 \ \mu_\mathrm{B}$/f.u., which is much
less than the Sm$^{3+}$ free ion value of $M_\mathrm{sat} =
g_{J}J\mu_\mathrm{B} = 0.71\ \mu_\mathrm{B}$/f.u.  This behavior
was attributed to screening of the magnetic moment due to the
Kondo effect. Like \SmFeP{},  the compound \SmFeSb{} has also been
found to be ferromagnetic; however, unlike the magnetization of
\SmFeP{}, \SmFeSb{} is characterized by a saturation magnetization
at $5$ K of $M_\mathrm{sat} = 0.7\ \mu_\mathrm{B}$/f.u., which
agrees with the Sm$^{3+}$ free ion value. After taking into
account the contribution from the FeSb$_{3}$ polyanions, the
effective magnetic moment of \SmFeSb{} was found using Van Vleck's
formula to be $\mu_\mathrm{eff} = 1.6\pm 0.2\
\mu_\mathrm{B}$/f.u., in good agreement with the calculated value
of $\mu_\mathrm{eff} = 1.66\
\mu_\mathrm{B}$/f.u.\cite{Danebrock96}

This paper reports x-ray diffraction, specific heat,
magnetization, and electrical resistivity measurements on the
compound \SmOsSb{}. These measurements revealed the presence of
crystalline electric field effects, heavy fermion behavior, and
magnetic order. Analysis of the magnetization using scaling theory
and modified Arrott plots shows signs of a ferromagnetic component
to the magnetic order.
%To gain a better understanding of these observed phenomena,
%crystalline electric field fits, scaling, and modified Arrott
%plots were employed in the analysis of the data.

\section{Experimental Details}

Single crystals of \SmOsSb{} were grown in a molten Sb flux.
Stoichiometric amounts of Sm (CERAC $99.9\%$) and Os (Colonial
Metals $99.95\%$) were combined with an excess of Sb (CERAC
$99.999\%$) in the atomic ratio $1:4:20$ and placed in a carbon
coated quartz tube, which was then evacuated and filled with $150$
Torr Ar prior to being sealed. The sealed tube was placed in a box
furnace, heated to $1050\ ^{\circ}$C for $48$ hours, and then
cooled to $700\ ^{\circ}$C at $2\ ^{\circ}$C/hr. The resulting
crystals were cubic and tended to form in large interconnected
clusters, with most of them much less than $1$ mm in dimension. A
small batch of ultra-high purity single crystals were also
produced using Sm (AMES $99.99\%$), Os (Alfa Aesar $99.999\%$),
and Sb (CERAC $99.999\%$). However, these single crystals were
only used as a subsequent check for impurities.

The quality of the single crystals was determined by x-ray powder
diffraction measurements, which were performed with a Rigaku D/MAX
B x-ray machine on a powder prepared by grinding several single
crystals.  Single crystal structural analysis was performed on two
single crystals with similar dimensions. Inspection with an
AXS-Gadds texture goniometer assured the high quality of the
specimens prior to x-ray intensity data collection. For the data
collection, a four-circle Nonius Kappa diffractometer equipped
with a CCD area detector employing graphite monochromated Mo
K$_{\alpha}$ radiation ($\lambda$ = 0.071073\,nm) was used.
Orientation matrix and unit cell parameters were derived using the
DENZO program.\cite{DENZO} No absorption corrections were
necessary because of the regular crystal shape and small
dimensions of the investigated crystals. The structure was refined
with the aid of the SHELXS-97 program.\cite{SHELXS97}

Specific heat $C(T)$ measurements were made between $0.6$ K and
$70$ K in a semi-adiabatic $^{3}$He calorimeter using a standard
heat pulse technique.  Many single crystals were combined for a
total mass of $49.34$ mg which were attached to the thermometer
(Cernox)/heater sapphire platform with about $7.44$ mg of Apiezon
N grease. The electrical resistivity $\rho$(T) of several samples
with dimensions of $\sim 1 \times{} 0.5 \times{} 0.5$ mm was
measured using a four probe technique from $1.1$ to $300$ K.
Magnetoresistance $\rho$(H,T) measurements were made using a four
probe ac technique in fields up to $9$ T in the $2$ to $300$ K
temperature range in a Quantum Design Physical Properties
Measurement System and in magnetic fields up to $18$ T down to
$\sim 0.05$ K using $^{3}$He-$^{4}$He dilution refrigerators at
UCSD ($0 - 8$ T) and the National High Magnetic Field Laboratory
at Los Alamos National Laboratory ($8 - 18$ T). All of the
resistivity measurements were made with a constant current of
\mbox{$100-300$ $\mu$A,} perpendicular to the applied magnetic
field. Magnetization measurements were made with Quantum Design
SQUID magnetometers in the temperature range $1.7$ to $300$ K in
magnetic fields up to $5$ T. Measurements of the dc and ac
magnetic susceptibility $\chi_\mathrm{dc}(T)$ and
$\chi_\mathrm{ac}(T)$, respectively, and the isothermal
magnetization $M(H)$ were made on a $5.37$ mg single crystal with
dimensions $\sim 0.6 \times{} 1.16 \times{} 1.18$ mm mounted such
that the field was applied along the long axis of the crystal.

\section{\label{sec:level}Results}

\subsection{Single crystal structural refinement}

Analysis of the x-ray powder diffraction pattern indicated single
phase \SmOsSb{} with a minor impurity peak of \OsSb{} $(\lesssim 5
\%)$. Table~\ref{StructureTBL} lists the results of a structural
refinement performed on x-ray diffraction data taken from two
\SmOsSb{} single crystals. The structural refinement indicated
that \SmOsSb{} has a \LaFeP-type structure,\cite{Jeitschko77} with
a lattice parameter $a = 9.3085$ \AA{} and a unit cell volume of
$806.6$ \AA$^{3}$.  The Sm site was found to be fully occupied and
Sm was found to have large isotropic thermal displacement
parameters $U_{ii}$ relative to those of Os and Sb.  The values of
$U_{ii}$ for Sm are consistent with the``rattling" behavior of the
filled skutterudite compounds.\cite{Sales96}

\subsection{\label{sec:level2}Specific Heat}

\begin{figure}[tbp]
    \begin{center}
    \includegraphics[width=3.375in]{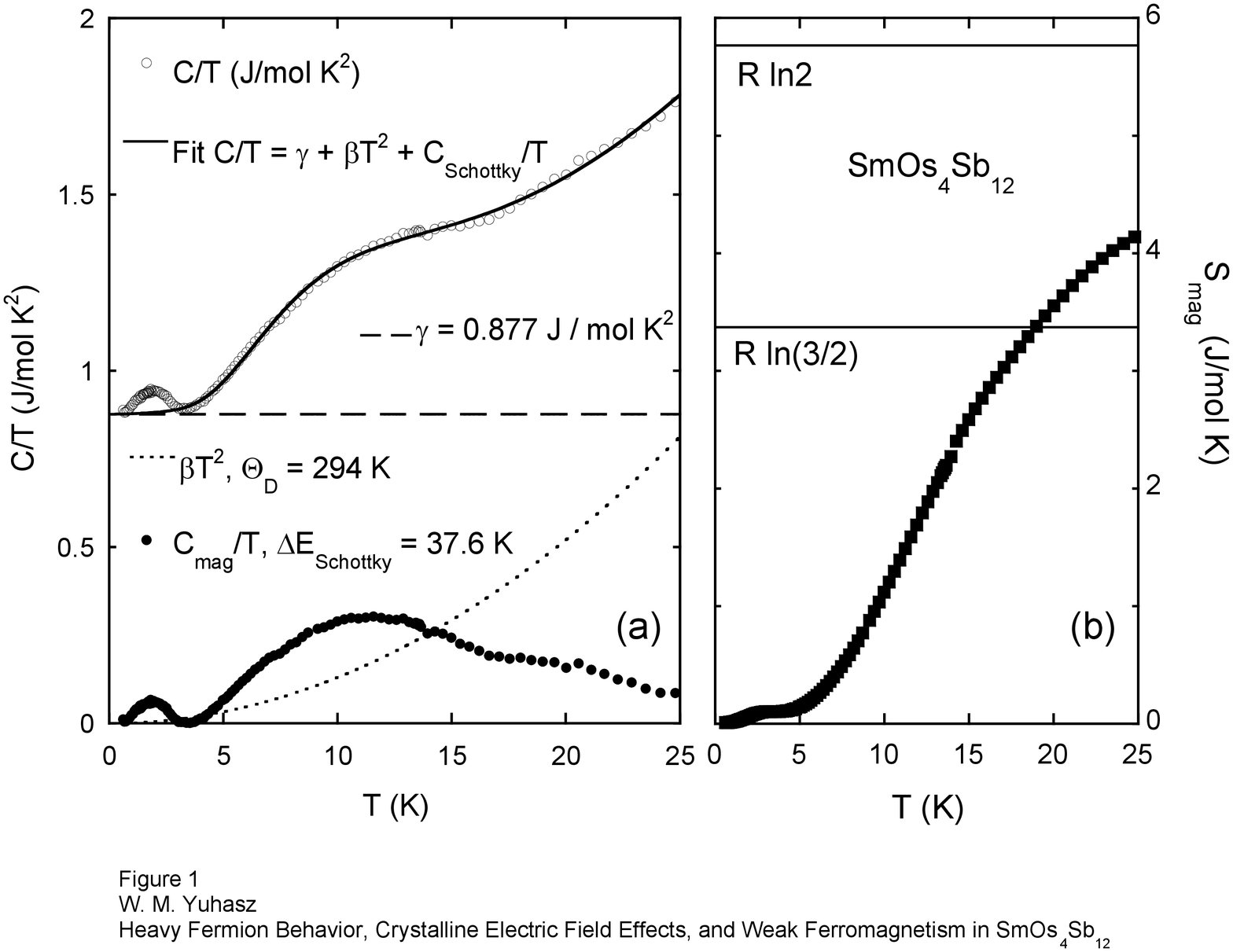}
    \end{center}
    \caption{(a) Plot of $C/T$ vs $T$ (open circles) and a fit to the data
    (solid line) that includes an electronic term (long dashed line), a lattice
    contribution (short dashed line), and a Schottky anomaly (see text for details).
    Also displayed is the magnetic contribution $C_\mathrm{mag}/T$ (filled circles),
    determined by subtracting the electronic and lattice terms from $C/T$.
    (b) Magnetic entropy $S_\mathrm{mag}$ of \SmOsSb{} as a function of $T$.}
    \label{SpecificHeat}
\end{figure}

Displayed in Fig.\ \ref{SpecificHeat}a is a plot of the specific
heat divided by temperature $C/T$ vs $T$ of \SmOsSb{} between
$0.6$ and $25$ K. Two features are readily apparent in the data
represented by the open circles in Fig.\ \ref{SpecificHeat}a.  The
first feature, a hump which peaks at $\sim 2$ K, can be associated
with the onset of ferromagnetic order when corroborated by
$\chi_\mathrm{ac}(T)$ and $M(H)$  measurements described later in
this article.  The ferromagnetic ordering temperature (Curie
temperature) was approximated by taking the maximum negative slope
of $C/T$, which occurs at $2.6$ K.  This feature is not as well
defined as that expected for a typical ferromagnetic transition
and may be due to the presence of an impurity phase.  However,
this hump is similar to a broad feature in the specific heat of
\PrFeSb{} that is associated with the occurrence of magnetic
order.\cite{Bauer02b}

The second feature observed in the specific heat data is a
Schottky-like anomaly centered at $\sim 10$ K. The data between
$3.5$ and $20$ K were fitted by an equation containing electronic,
lattice, and Schottky terms. These results are also displayed in
Fig.\ \ref{SpecificHeat}a. The Debye temperature
$\Theta_\mathrm{D} \approx 294$ K inferred from the fit is typical
for a filled skutterudite compound, while the resultant electronic
specific heat coefficient $\gamma \approx 880$ mJ/mol K$^{2}$ is
extremely large.  It is important to note that this value of
$\gamma$ is fairly insensitive to the Debye temperature
$\Theta_\mathrm{D}$ and the splitting $\Delta{}E$ of the crystal
field levels, changing by less than $5\%$ for various reasonable
values of $\Theta_\mathrm{D}$ and $\Delta{}E$. The effective mass
$m^{*}$ can be estimated from $\gamma$ using the relation
\begin{equation}\label{mstar}
\gamma =
\frac{\pi^{2}(Z/\Omega)k_\mathrm{B}^{2}m^{*}}
{\hbar^{2}k_\mathrm{F}^{2}},
\end{equation}
where $Z$ is the number of charge carriers per unit cell, $\Omega$
is the unit cell volume, and $k_\mathrm{F} =
(3\pi^{2}Z/\Omega)^{1/3}$ is the Fermi wave vector using a
spherical Fermi surface approximation. We assume that $Z = 2$,
since there are two formula units per unit cell and in each
formula unit, Sm$^{3+}$ contributes $3$ electrons and each of the
(OsSb$_{3}$)$^{-1}$ polyanions contributes one hole.  Eq.
\ref{mstar} yields $m^{*} \approx 170\ m_\mathrm{e}$, where
$m_\mathrm{e}$ is the free electron mass, revealing that \SmOsSb{}
is a heavy fermion compound.

The magnetic contribution to the entropy, $S_\mathrm{mag}$, shown
in Fig.\ \ref{SpecificHeat}b was determined by subtracting the
electronic and lattice contributions from $C/T$, extrapolating to
zero temperature, and then integrating over $T$. At the $\sim 2.6$
K magnetic transition, the magnetic entropy was found to be $\sim
94$ mJ/mol K ($1.6\%$ of $R \ln 2$) and it only reaches a value of
$\sim 4.2$ J/mol K by $25$ K, where the Schottky contribution is
in its high temperature tail and significantly reduced.

\subsection{\label{sec:level3}Electrical Resistivity}

\begin{figure}[tbp]
    \begin{center}
    \includegraphics[width=3.375in]{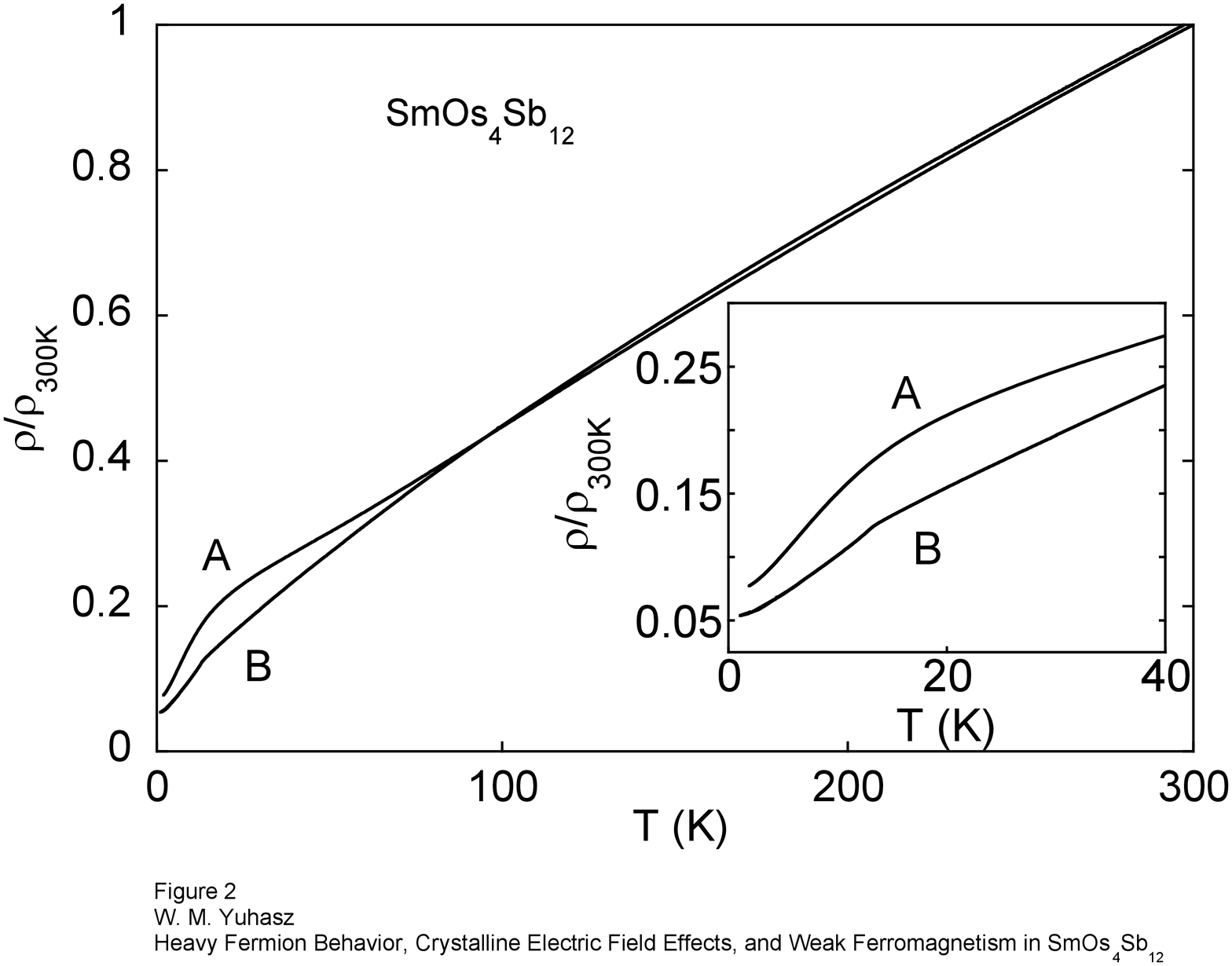}
    \end{center}
    \caption{Electrical resistivity $\rho$ vs temperature $T$ for \SmOsSb{} samples A and B.
    The low temperature behavior is shown in the inset.} \label{Resistivity}
\end{figure}

Electrical resistivity measurements were performed on several
single crystals of \SmOsSb{}.  Figure\ \ref{Resistivity} displays
the room temperature normalized electrical resistivity
\mbox{$\rho/\rho(294$ K)} vs $T$ for two representative samples.
Samples A and B both exhibit metallic behavior with residual
resistivity ratio (RRR) values of $12.7$ and $19.1$, respectively,
and room temperature resistivity $\rho(294$ K) values of $313$ and
$379$ $\mu\Omega$ cm, respectively. At low temperatures, there is
a broad hump in $\rho(T)$ between $6$ K and $40$ K for sample A
and a kink in $\rho(T)$ at $\sim 12.5$ K for sample B. These
features occur well above $T_\mathrm{mag} \approx 2.6$ K
determined from $C(T)$, as well as from $\chi_\mathrm{ac}(T)$ and
$M(H,T)$ data presented later.  The sample dependence of the
electrical resistivity along with the high room temperature
resistivity values indicate the presence of atomic disorder.

\begin{figure}[tbp]
    \begin{center}
    \includegraphics[width=3.375in]{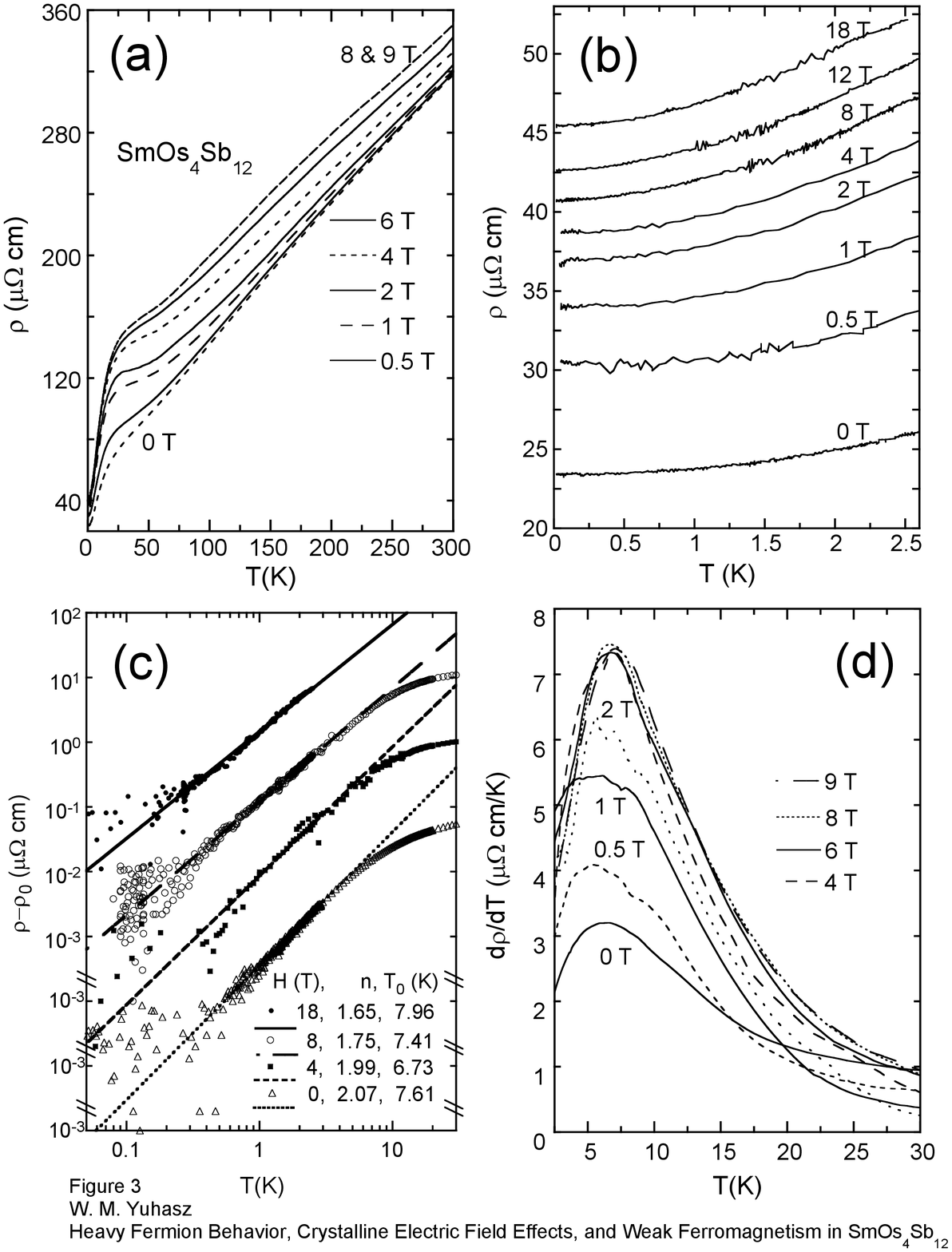}
    \end{center}
    \caption{Electrical resistivity $\rho(T)$ at various fields for \SmOsSb{}
    (Sample A).  (a) $\rho(T,H)$ for \mbox{$2$ K $\leq T \leq$ $300$ K} and $0$ T $\leq H
    \leq$ $9$ T. (b) $\rho(T,H)$ for $0.02$ K $\leq T \leq$ $2.5$ K and \mbox{$0$ T $\leq H
    \leq$ $18$ T}. (c) $\rho-\rho_{0}$ vs T on a log-log scale. The lines are power law fits
    to the data of the form \mbox{$\rho = \rho_{0}[1 + (T/T_{0})^{n}]$}.
    (d) ${\it d}\rho$/${\it d}T$ vs $T$ at fields up to $9$ T showing a peak at
    $\sim 7.5$ K for all fields.}
    \label{SmOsSbrhoTH}
\end{figure}

\begin{figure}[tbp]
    \begin{center}
    \includegraphics[width=3.375in]{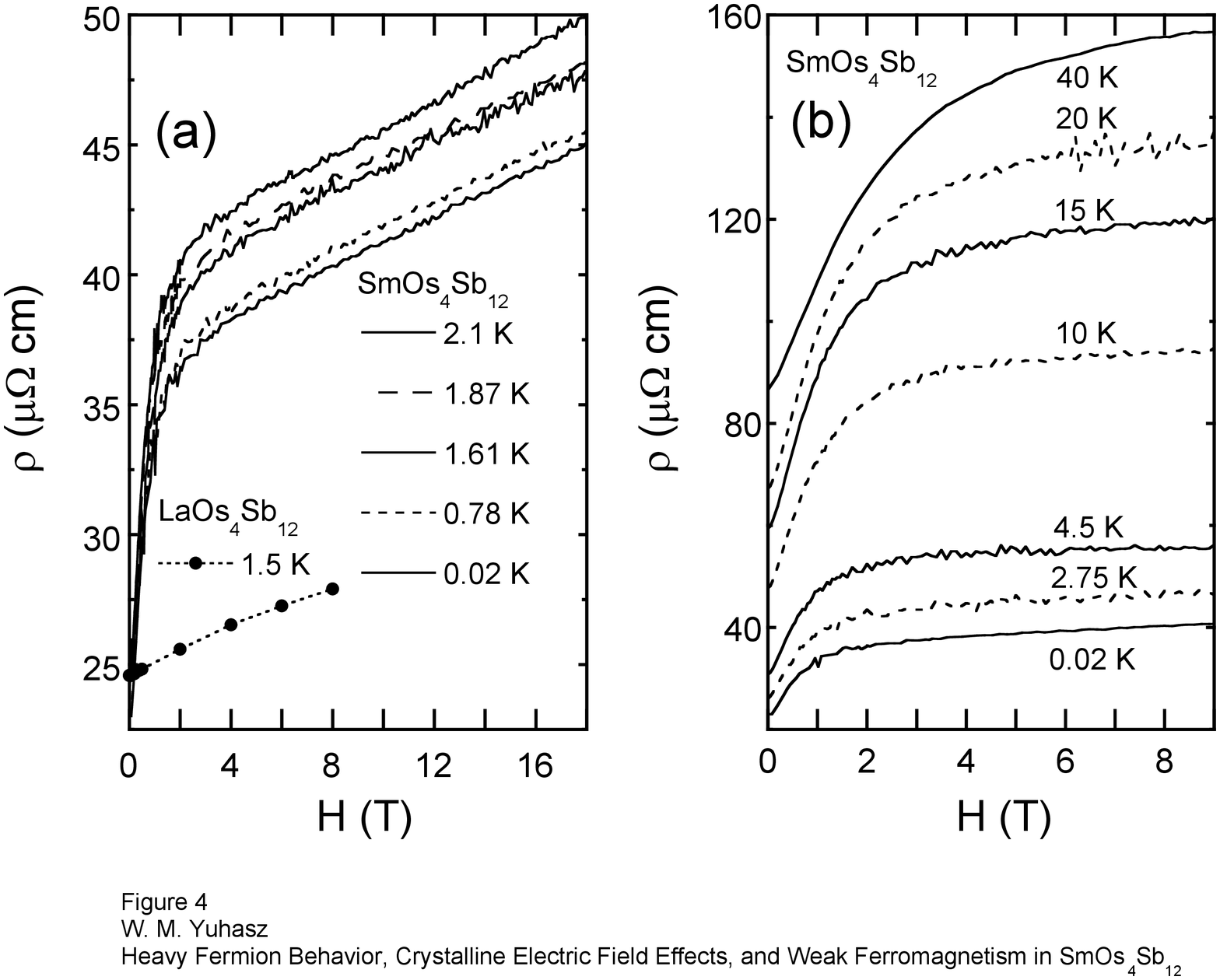}
    \end{center}
    \caption{Electrical resistivity $\rho(H)$ at various temperatures for \SmOsSb{}
    (Sample A) along with $\rho(H)$ at $1.5$ K for \LaOsSb{}.  (a) $\rho(H,T)$
    for $0$ T $\leq H \leq$ $9$ T and $0.02$ K $\leq T \leq$ $40$ K. (b) $\rho(H,T)$
    for $0$ T $\leq H \leq$ $18$ T and $0.02$ K $\leq T \leq$ $2.75$ K. The $\rho(H)$
    data for \LaOsSb{} (solid circles) at \mbox{$1.5$ K} (shifted upwards by adding
    $20$ $\mu\Omega$ cm) for comparison with the low temperature $\rho(H)$ data of
    \SmOsSb{}.}
    \label{SmOsSbrhoHT}
\end{figure}

The temperature and magnetic field dependencies of the electrical
resistivity for \SmOsSb{} are shown in Figs. \ref{SmOsSbrhoTH} and
\ref{SmOsSbrhoHT}.  Figure \ref{SmOsSbrhoTH}a displays the
electrical resistivity of \SmOsSb{} from $2$ to $300$ K in fields
up to $9$ T.  In general, the resistivity increases with
increasing field at all temperatures. The shoulder in $\rho(T)$,
which is observed at $\sim 30$ K, becomes more prominent as the
magnetic field increases. The behavior of the low temperature
$\rho(H)$ data above \mbox{$4$ T} for \SmOsSb{} is very similar to
that of \LaOsSb{}; however, below $4$ T, the $\rho(H)$ data
exhibit a rapid increase with $H$ whose origin is not understood.

The heavy fermion behavior inferred from the specific heat
measurements is also reflected in the electrical resistivity.  The
electrical resistivity of a typical {\it f}-electron heavy fermion
compound has a relatively weak temperature dependence at high
temperatures and then decreases rapidly with decreasing
temperature below a characteristic ``coherence temperature'',
until, at the lowest temperatures, the resistivity varies as
$T^{2}$. The $T^{2}$ dependence is indicative of Fermi-liquid
behavior and is strong enough to be readily observable in most
heavy fermion compounds. In order to ascertain whether the
resistivity of \SmOsSb{} follows this $T^{2}$ dependence at low
temperatures, power law fits of the form $\rho = \rho_{0}[1 +
(T/T_{0})^{n}]$ (where $\rho_{0}$ is the residual resistivity and
${T_0}$ is a characteristic temperature) were made to the
$\rho(T)$ data in the temperature range from $\sim 0.05$ to $\sim
10$ K up to $8$ T and from $\sim 0.02$ to $2.6$ K from $10$ to
$18$ T (Fig.\ \ref{SmOsSbrhoTH}c). The fits show that the exponent
$n$ is approximately $2$ up to $4$ T, consistent with Fermi-liquid
behavior, and then decreases with increasing field, possibly
indicating a cross over to field-induced non-Fermi-liquid
behavior. The value of $T_{0}$ determined from the power law fits
varies between $\sim 7$ and $8$ K, and may be associated with the
peak observed in the derivative of the resistivity ${\it d}
\rho$/${\it d} T$ vs $T$, shown in Figure \ref{SmOsSbrhoTH}d,
which occurs around \mbox{$7.5$ K} at all fields between $0$ and
$9$ T. This low value of the scaling temperature $T_{0}$ is
consistent with the large value of $\gamma$.

The coefficient $A$ $( = \rho_{0}/( T_{0}^{2} ))$ of the $T^{2}$
term in the electrical resistivity is often found to follow the
Kadowaki-Woods (KW) relation $A$/$\gamma^2 = 1 \times{} 10^{-5}
\mu\Omega$ cm mol$^{2}$ K$^{2}$ mJ$^{-2}$ (where $\gamma$ is the
electronic specific heat coefficient).\cite{Kadowaki86} For
\SmOsSb{}, the power law fits to the electrical resistivity result
in an $A$/$\gamma^2$ ratio of $\sim 0.5 \times{} 10^{-6}
\mu\Omega$ cm mol$^{2}$ K$^{2}$ mJ$^{-2}$ at $0$ T, which is $\sim
20$ times smaller than the value expected from the KW relation.
However, Tsujii {\it et al.} recently found a different empirical
relation, $A$/$\gamma^2 \approx 0.4 \times{} 10^{-6} \mu\Omega$ cm
mol$^{2}$ K$^{2}$ mJ$^{-2}$, from studies of several Yb-based
compounds (such as YbCu$_{5}$, YbAl$_{3}$, and YbInCu$_{4}$), some
Ce-based compounds (CeNi$_{9}$Si$_{4}$ and CeSn$_{3}$), and some
transition metals (such as Fe, Pd, and Os).\cite{Tsujii03}  To
explain this new relation, Tsujii {\it et al.} suggested a
relation to the ground state degeneracy of the system, which has
been developed further by Kontani in the form of a generalized
Kadowaki-Woods relation.\cite{Tsujii03,Kontani04}  The
$A$/$\gamma^2$ value for \SmOsSb{} seems to be in much better
agreement with this new relation and, thereby, consistent with the
behavior of several other heavy fermion compounds.

\subsection{\label{sec:level4}Magnetization and Magnetic Susceptibility}

\begin{figure}[tbp]
    \begin{center}
    \includegraphics[width=3.375in]{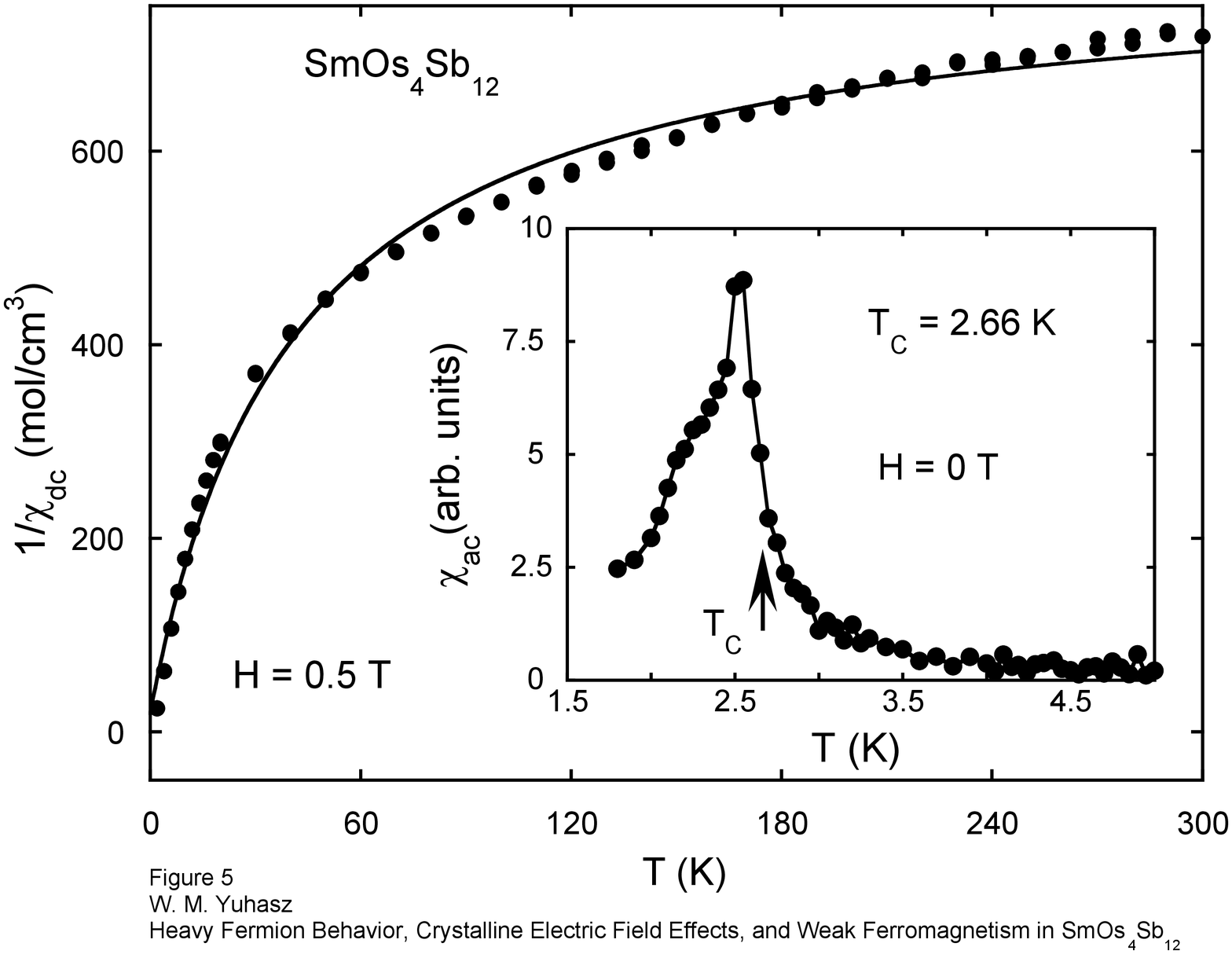}
    \end{center}
    \caption{Inverse dc magnetic susceptibility $\chi_\mathrm{dc}^{-1}$, measured at
    $0.5$ T, vs temperature $T$ (filled circles) for \SmOsSb{}. The line represents
    a fit of a Curie-Weiss law with a temperature independent Van Vleck term
    using Eq. \ref{chiequation} to the $\chi_\mathrm{dc}^{-1}(T)$ data. The inset shows the
    temperature dependence of the ac magnetic susceptibility $\chi_\mathrm{ac}$.
    The arrow denotes the $T_\mathrm{C}$ of \SmOsSb{} (defined as the midpoint of the
    transition on the paramagnetic side).}
    \label{X(T)}
\end{figure}

Inverse dc magnetic susceptibility $\chi_\mathrm{dc}^{-1}$ vs $T$
data for \SmOsSb{} are shown in Fig.\ \ref{X(T)}. Since Sm$^{3+}$
ions have relatively low-energy angular momentum states above the
Hund's rule $J = 5/2$ ground state, a simple Curie-Weiss law was
unable to describe the data. Previous work \cite{Hamaker79} has
shown that $\chi(T)$ for Sm compounds can often be reasonably well
described without considering CEF splitting by the equation:
\begin{equation}\label{chiequation}
\chi(T)  = \left(\frac
{N_\mathrm{A}}{k_\mathrm{B}}\right)\left[\frac
{\mu_\mathrm{eff}^{2}}{3(T - \theta_\mathrm{CW})} + \frac
{\mu_{B}^{2}}{\delta}\right],
\end{equation}
where $N_\mathrm{A}$ is Avogadro's number, $\mu_\mathrm{eff}$ is
the effective magnetic moment, $\theta_\mathrm{CW}$ is the
Curie-Weiss temperature, $\mu_\mathrm{B}$ is the Bohr magneton,
and $\delta = 7\Delta{}/20$, where $\Delta$ is the energy
(expressed in units of K) between the Hund's rule $J = 5/2$ ground
state and the $J = 7/2$ first excited state. Equation
\ref{chiequation} consists of a Curie-Weiss term due to the $J =
5/2$ ground state contribution and a temperature independent Van
Vleck term due to coupling with the first excited $J = 7/2$
multiplet.  The theoretical Sm$^{3+}$ free ion moment is
\mbox{$\mu_\mathrm{eff} = g_{J}(J(J+1))^{1/2}\mu_\mathrm{B} =
0.845\ \mu_\mathrm{B}$/f.u.}, where $g_{J} = 0.286$ is the
Land\'{e} g-factor and $J = 5/2$. The best overall fit of Eq.
\ref{chiequation} to the $\chi_\mathrm{dc}^{-1}(T)$ data, shown in
Fig.\ \ref{X(T)}, results in the parameters $\theta_\mathrm{CW} =
-0.99$ K, $\delta = 300$ K, and $\mu_\mathrm{eff} = 0.63\
\mu_\mathrm{B}$/f.u. The value of $\mu_\mathrm{eff} = 0.63\
\mu_\mathrm{B}$/f.u. is somewhat less than the theoretical
Sm$^{3+}$ free ion value of $\mu_\mathrm{eff} = 0.845\
\mu_\mathrm{B}$/f.u., while $\delta = 300$ K yields $\Delta{} =
20\delta/7 = 850$ K, which is much less than the $\Delta{} \sim
1500$ K value estimated for free Sm$^{3+}$ ions.\cite{Van Vleck32}
However, low values of $\Delta$ have previously been inferred from
the fits to $\chi_\mathrm{dc}(T)$ data for other Sm-based
compounds such as SmRh$_{4}$B$_{4}$.\cite{Hamaker79}

\begin{figure}[tbp]
    \begin{center}
    \includegraphics[width=3.375in]{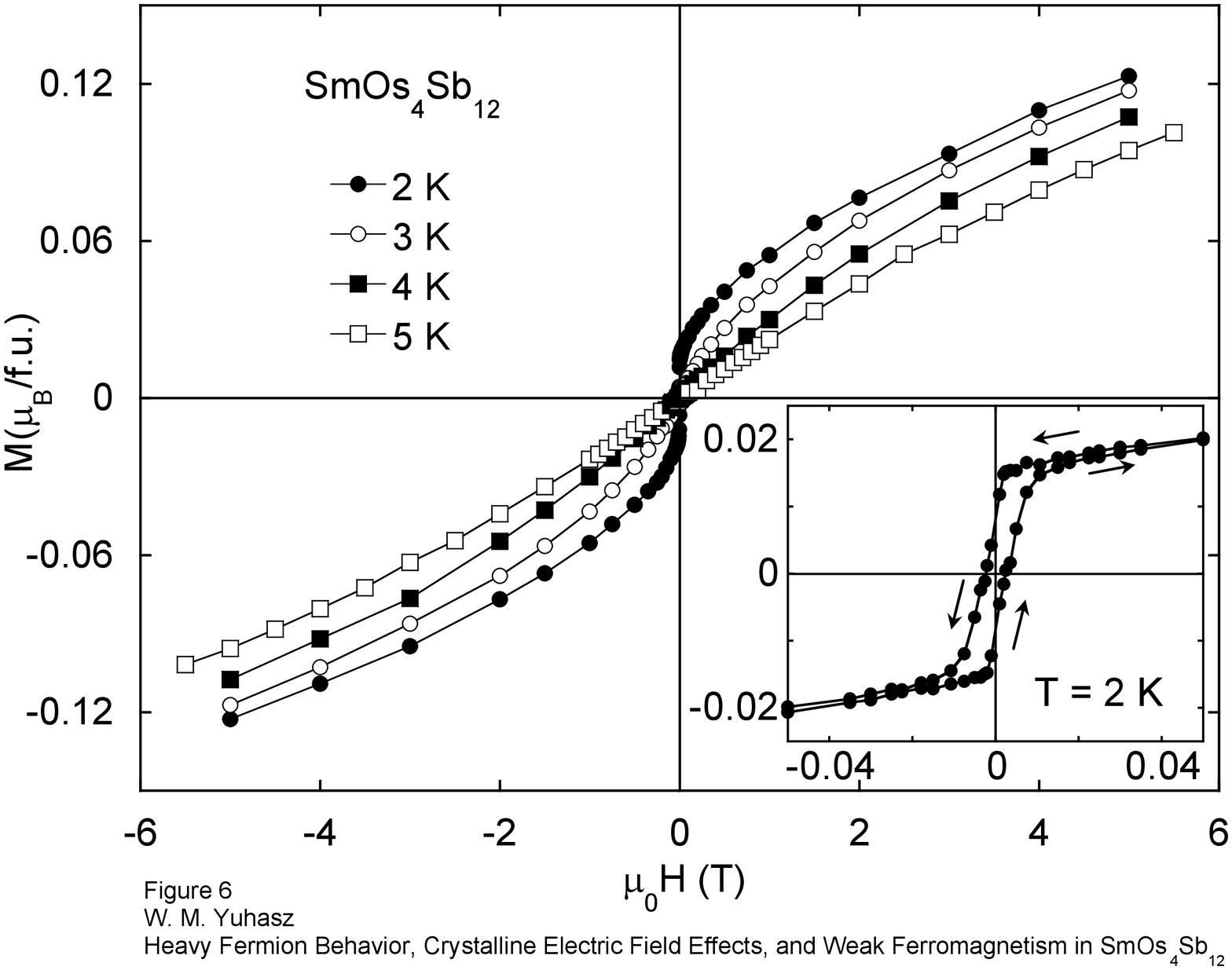}
    \end{center}
    \caption{$M(H)$ isotherms between $-5$ T and $5$ T for \SmOsSb{} at
    $2$ K, $3$ K, $4$ K, and $5$ K.  The inset displays the low field
    behavior of the $2$ K isotherm.}
    \label{M(H)}
\end{figure}

The magnetic properties of \SmOsSb{} were also characterized by
measuring $\chi_\mathrm{ac}(T)$ and $M(H,T)$ at low temperatures.
The $\chi_\mathrm{ac}( T)$ data (Fig.\ \ref{X(T)}, inset) exhibit
a peak indicative of a magnetic transition at $T_\mathrm{C} =
2.66$ K, where $T_\mathrm{C}$ is defined as the temperature of the
midpoint of the change in $\chi_\mathrm{ac}$ on the paramagnetic
side. The results of isothermal $M(H)$ measurements, made in the
vicinity of the transition, are shown in Fig.\ \ref{M(H)}.  The
magnetic transition can clearly be seen in the $M(H)$ isotherms
where the approximately linear behavior at $5$ K becomes nonlinear
at lower temperatures and hysteretic at $2$ K (Fig.\ \ref{M(H)},
inset). These results reveal that some type of magnetic order with
a weak ferromagnetic component occurs below $2.66$ K. At $2$ K,
the remnant magnetization $M_\mathrm{R}$ is $\sim 0.015(1)
\mu_\mathrm{B}$/f.u. and the coercive field $H_\mathrm{C}$ is
$\sim 2.5(1) \times{} 10^{-3}$ T. Even though saturation is not
achieved at $2$ K, which is only \mbox{$\sim 0.5$ K} away from
$T_\mathrm{C}$, a magnetization value $M$ of $\sim 0.122(1)
\mu_\mathrm{B}$/f.u. is obtained at $5$ T which is only $17\%$ of
the theoretical value of $M_\mathrm{sat} = g_{J}J\mu_\mathrm{B} =
0.71\ \mu_\mathrm{B}$/f.u.

\begin{figure}[tbp]
    \begin{center}
    \includegraphics[width=3.375in]{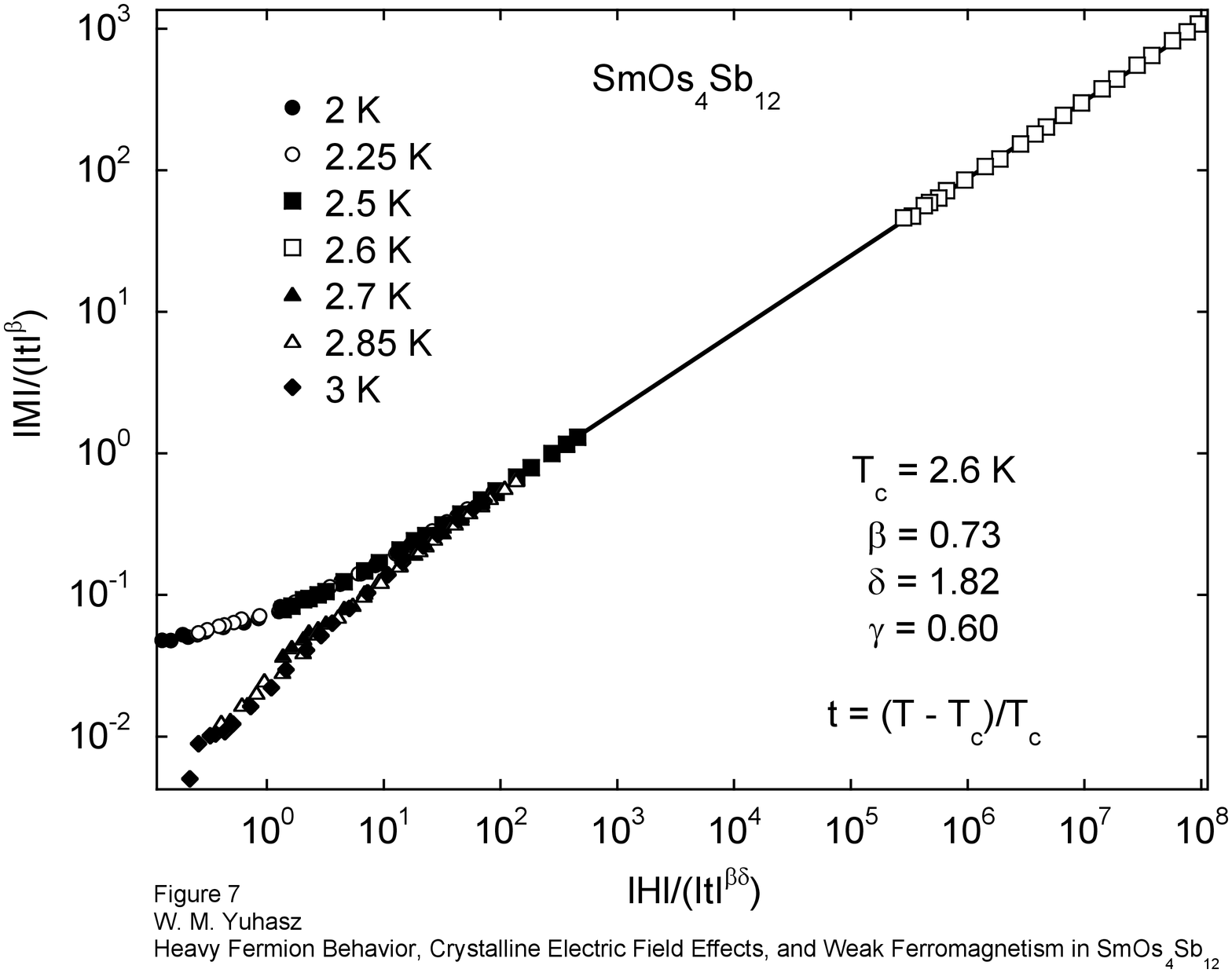}
    \end{center}
    \caption{Scaling plot $|M|/|t|^{\beta}$ vs $|H|/|t|^{\beta\delta}$ for
    \SmOsSb{}.  The line is a guide to the eye.}
    \label{Scaling}
\end{figure}

\begin{figure}[tbp]
    \begin{center}
    \includegraphics[width=3.375in]{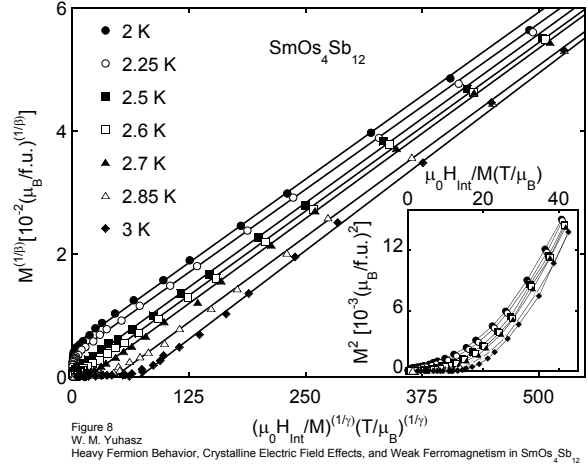}
    \end{center}
    \caption{Modified Arrott plot for \SmOsSb. The inset shows the
     conventional Arrott plot for \SmOsSb{} using the the critical exponents from
     the molecular field approximation ($\beta = 1/2$ and $\gamma = 1$). }
     \label{Arrott}
\end{figure}

Arrott plots were constructed in an attempt to determine the Curie
temperature $T_\mathrm{C}$, the spontaneous magnetization
$M_\mathrm{S}$, and the initial susceptibility $\chi_{0}$.  An
Arrott plot consists of $M^{2}$ vs $(H/M)$ isotherms, where $M$ is
magnetization and $H$ is the internal field. In general, the
$M^{2}$ vs $(H/M)$ isotherms form a series of lines for a
ferromagnetic compound that are parallel near $T_\mathrm{C}$,
where $T_\mathrm{C}$ corresponds to the isotherm that passes
through the origin. However, in the case of \SmOsSb{}, the $M^{2}$
vs $(H/M)$ isotherms are strongly curved, as shown in the inset to
Fig.\ \ref{Arrott}.  To overcome this difficulty, the $M(H,T)$
data were analyzed using a modified Arrott plot, $M^{1/\beta}$ vs
$(H/M)^{1/\gamma}$ (where $\beta$ and $\gamma$ are critical
exponents), which is based on the Arrott-Noakes equation of
state.\cite{Arrott67}  To construct the modified Arrott plot, it
was necessary to determine the critical exponents $\beta$ and
$\gamma$.  This was accomplished by estimating the value of the
critical exponent $\delta$ using the relation $M \sim H^{1/\delta}
(T=T_\mathrm{C})$ and plotting {\it d}ln$(\mu_{0}H)/${\it
d}ln$(M)$ vs $\mu_{0}H$ for all the measured isotherms. The
isotherm with the slope closest to zero was found to be the one
for $T = 2.6$ K, and the average value for {\it
d}ln$(\mu_{0}H)/${\it d}ln$(M)$ above $\mu_{0}H = 0.05$ T for this
isotherm gives $\delta = 1.82(5)$. These values were then analyzed
using scaling theory (Fig.\ \ref{Scaling}), from which
$|M|/|t|^{\beta}$ is plotted as a function of
$|H|/|t|^{\beta\delta}$ where $t = (T-T_\mathrm{C})/T_\mathrm{C}$;
on this plot, the isotherms collapse onto two universal curves
with the isotherms for $T > T_\mathrm{C}$ on one branch and those
for $T < T_\mathrm{C}$ on the other. Based on this scaling
analysis, values of $\beta = 0.73(5)$ and $\delta = 1.82(5)$ were
determined, while $\gamma = 0.60(5)$ was obtained from the Widom
scaling relation $\delta = 1 + \gamma/\beta$.\cite{Binney92} The
resulting value of $T_\mathrm{C}$ from the scaling analysis is
$2.60(5)$ K, which agrees well with $T_\mathrm{C}$ determined from
$\chi_\mathrm{ac}(T)$ measurements on the same crystal.

\begin{figure}[tbp]
    \begin{center}
    \includegraphics[width=3.375in]{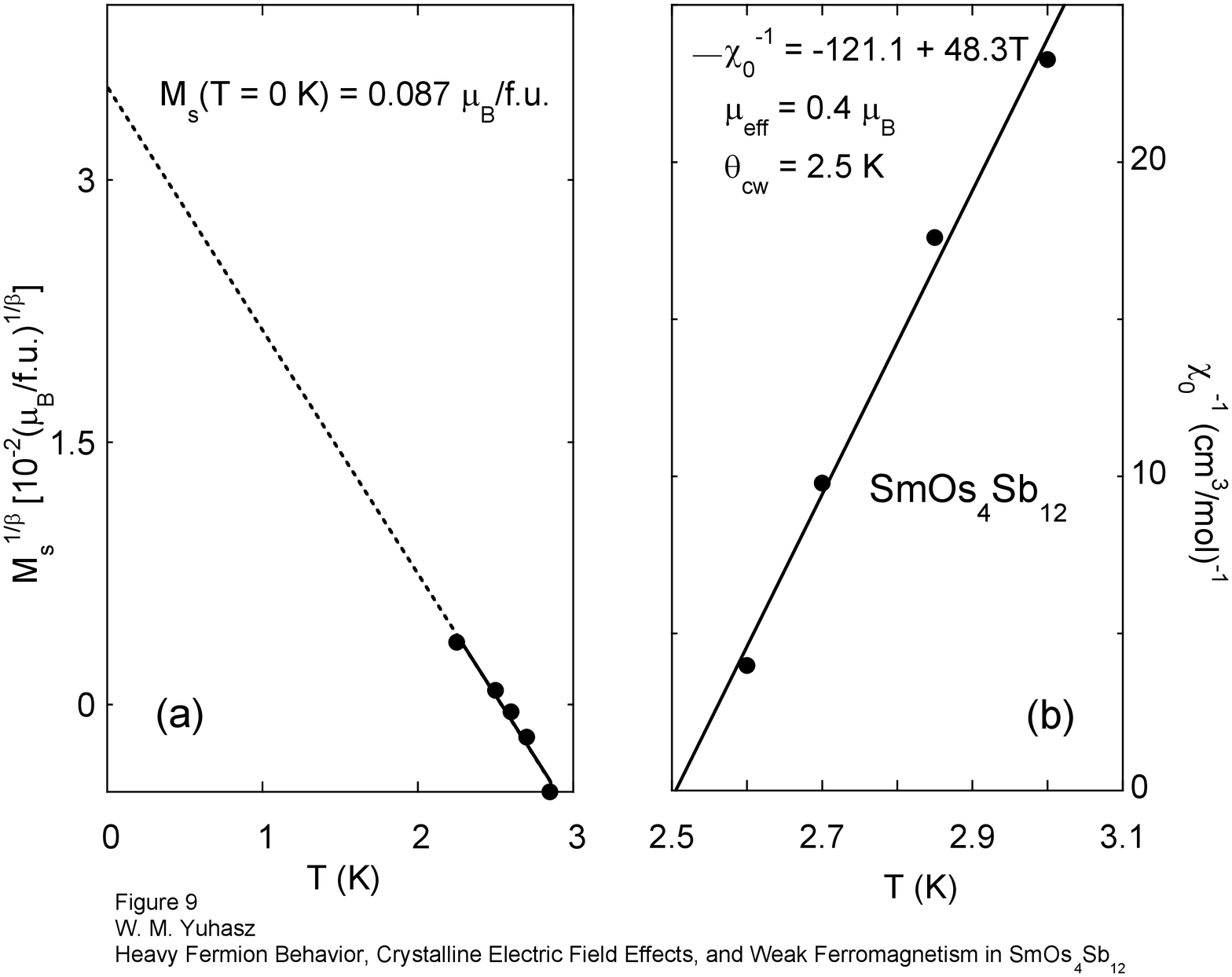}
    \end{center}
    \caption{(a)  Spontaneous magnetization $M_\mathrm{S}^{1/\beta}$ vs $T$
    along with a linear fit (solid line) and extrapolation to $0$ K (dashed line).
    (b) Inverse initial susceptibility $\chi_{0}(T)^{-1}$ vs $T$ along with
    a Curie-Weiss fit (line).}
    \label{MsXo}
\end{figure}

The critical exponents determined from the scaling analysis were
then used to construct the modified Arrott plots, shown in Fig.\
\ref{Arrott}. With the correct critical exponents for \SmOsSb{},
the isotherms in the modified Arrott plot were linear and parallel
close to $T_\mathrm{C}$ in the high field region from $0.15$ T to
$3$ T. Linear fits to the $M^{1/\beta}$  vs $(H/M)^{1/\gamma}$
data were made in this field range as shown in Fig.\ \ref{Arrott};
the intercepts of the fits were then used to determine the Curie
temperature $T_\mathrm{C}$, the initial susceptibility $\chi_{0}$,
and the spontaneous magnetization $M_\mathrm{S}$ for each
isotherm.  The value of $T_\mathrm{C}$ from the modified Arrott
plots agrees well with the scaling analysis result of
$T_\mathrm{C} = 2.60(5)$ K to within $\sim 0.1$ K.  A Curie-Weiss
fit of $\chi_{0}^{-1}(T)$ (Fig.\ \ref{MsXo}a) resulted in
$\theta_\mathrm{CW} = 2.5$ K and $\mu_\mathrm{eff} = 0.4\
\mu_\mathrm{B}$/f.u.  These values are not in agreement with the
earlier fit to the $\chi_\mathrm{dc}^{-1}(T)$ data using Eq.
\ref{chiequation} since the $\chi_\mathrm{dc}^{-1}(T)$ data were
fit from $2$ to $300$ K while the $\chi_{0}^{-1}(T)$ data were fit
only near $T_\mathrm{C}$.

\section{\label{sec:level}Discussion}

\subsection{Crystalline Electric Field Analysis}
In the presence of a cubic crystalline electric field (CEF), the
six-fold degenerate Sm$^{3+}$ \mbox{$J = 5/2$} multiplet splits
into a $\Gamma_{7}$ doublet and a $\Gamma_{8}$ quartet. Although
it has been shown that the filling atom (Sm) in the filled
skutterudites experiences tetrahedral CEF
splitting,\cite{Takegahara01} cubic CEF splitting was used instead
for simplicity. The best fit to the $C/T$ data for \SmOsSb{},
obtained by scaling the CEF Schottky contribution by $0.58$,
resulted in a $\Gamma_{7}$ ground state and a $\Gamma_{8}$ excited
state separated by $\Delta{}E \approx 38$ K. The scaling that was
necessary could imply that most of the entropy of the $4f$
electrons in \SmOsSb{} is associated with the Schottky-like
anomaly while the rest resides within the heavy quasiparticles and
magnetic ordering.

\begin{figure}[tbp]
    \begin{center}
    \includegraphics[width=3.375in]{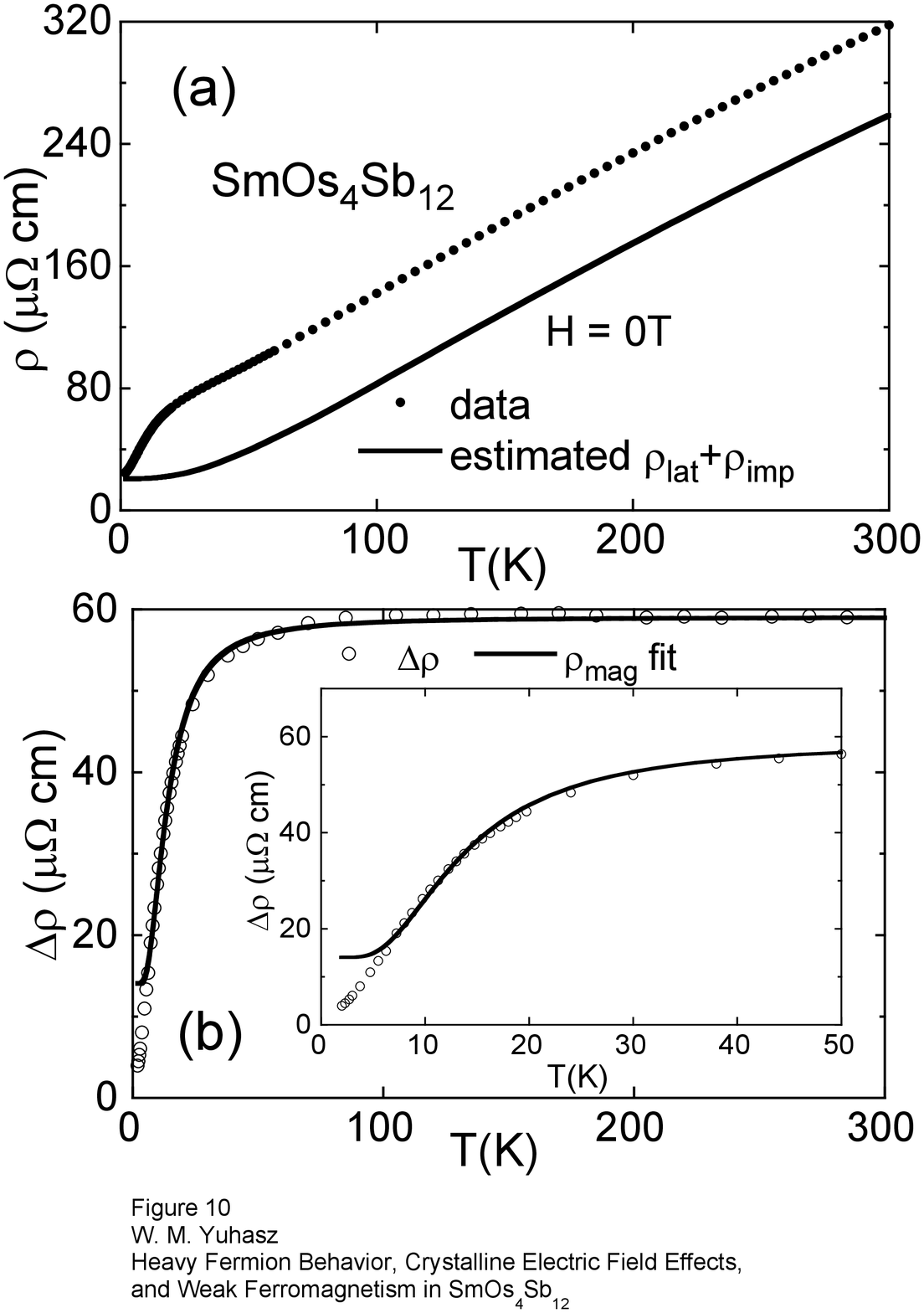}
    \end{center}
    \caption{(a) Zero-field electrical resistivity $\rho$ and the estimated
    $\rho_\mathrm{lat} + \rho_\mathrm{imp}$ vs $T$ for \SmOsSb{} (Sample A), where
    $\rho_\mathrm{imp} \sim 21$ $\mu\Omega$ cm. (b) $\Delta\rho$
    ($= \rho - \rho_\mathrm{lat} - \rho_\mathrm{imp}$) vs $T$ compared with a
    fit of the resistivity due to s-f exchange scattering of electrons from the $\Gamma_{7}$
    doublet ground state and the $\Gamma_{8}$ quartet excited state separated by
    $\sim 33$ K due to the CEF (solid line). The fit gives a good description of the
    $\Delta\rho(T)$ data from $7$ to $300$ K. The inset shows the quality of the fit at
    low temperatures.}
    \label{SmOsSbCEF0TFit}
\end{figure}

The zero-field electrical resistivity $\rho(T)$ of \SmOsSb{} is
shown in Fig.\ \ref{SmOsSbCEF0TFit}a (sample A). The resistivity
$\rho(T)$ has an approximately linear-$T$ dependence between $50$
and $300$ K and drops rapidly below $\sim 50$ K.  To determine if
the feature below $50$ K was due to CEF splitting of the Sm$^{3+}
J = 5/2$ multiplet, it was necessary to subtract a lattice
contribution $\rho_\mathrm{lat}$ and an impurity contribution
$\rho_\mathrm{imp}$ $(\sim 21$ $\mu\Omega$ cm) from the
resistivity data, yielding an incremental resistivity $\Delta\rho$
(where $\Delta\rho = \rho-\rho_\mathrm{lat}-\rho_\mathrm{imp}$).
Usually, $\rho_\mathrm{lat}$ is estimated from an isostructural
nonmagnetic reference compound; in the case of \SmOsSb{},
\LaOsSb{} was used. However, above 100 K, $\rho(T)$ of \LaOsSb{}
exhibits a significant amount of negative curvature, which is
common in La-based compounds (such as
LaAl$_{2}$).\cite{Slebarski85,Maple69} The curvature is generally
less pronounced in Y- and Lu-based compounds, which have empty and
filled $4f$-electron shells, respectively. However, since the
compounds \YOsSb{} and \LuOsSb{} have not yet, to our knowledge,
been prepared, an estimate of $\rho_\mathrm{lat}$ for \SmOsSb{}
was made from $2$ to $300$ K.  This estimate was derived by
shifting the linearly T-dependent resistivity of \SmOsSb{} above
$100$ K, where the T-dependence was assumed to be completely due
to electron-phonon scattering, such that it matched smoothly with
the resistivity of \LaOsSb{} below $100$ K. These data were then
combined to represent $\rho_\mathrm{lat}$ from $2$ to $300$ K in
Fig. \ref{SmOsSbCEF0TFit}a. After subtracting this estimated
lattice contribution, $\Delta\rho$ was plotted as shown in Fig.\
\ref{SmOsSbCEF0TFit}b and compared with the calculated resistivity
due solely to {\it s-f} exchange scattering ($\rho_\mathrm{mag}$)
from the Sm$^{3+}$ $4f$ energy levels in the CEF. The exchange
scattering contribution $\rho_\mathrm{mag}$ for \SmOsSb{} was
calculated for CEF splitting of the Hund's rule $J=5/2$ multiplet
for Sm$^{3+}$, similar to a procedure described elsewhere for
\PrOsSb{},\cite{Frederick03} which was based on work by Andersen
{\it et al.}.\cite{Andersen74}  Since Sm$^{3+}$ has a magnetic
ground state and, in order to simplify the analysis, a
contribution to the CEF resistivity due to aspherical Coulomb
scattering was not considered. The fit of the calculated
$\rho_\mathrm{mag}$(T) to the $\Delta\rho$(T) data was quite good,
except below $\sim 7$ K, where the discrepancy may be due to the
ferromagnetic phase transition that occurs at $\sim 2.6$ K or the
development of the coherent heavy Fermi liquid ground state. Based
on this fit, a splitting of $\sim 33$ K between the $\Gamma_{7}$
doublet ground state  and $\Gamma_{8}$ quartet excited state was
inferred. In general, the splitting that results from the CEF fit
of $\rho_\mathrm{mag}$(T) to the $\Delta\rho$(T) data is in
reasonable agreement with the value ($38$ K) determined from
specific heat measurements. However, the possibility that the drop
in the resistivity below $\sim 50$ K is due to the development of
the coherent heavy Fermi liquid, rather than CEF splitting of the
Sm$^{3+}$ $J=5/2$ multiplet, cannot be ruled out.

The fit of Eq.\ \ref{chiequation} to the
$\chi_\mathrm{dc}^{-1}(T)$ data yielded a good overall description
of the $\chi_\mathrm{dc}(T)$ data without incorporating CEF
effects. Fits that considered CEF effects in addition to the
splitting between the $J=5/2$ and $J=7/2$ multiplets (not shown)
did not vary significantly from the results of Eq.\
\ref{chiequation}. Following the modified Arrott plot analysis, it
was also found that the value of $\mu_\mathrm{eff}$ determined
from the fit to the $\chi_{0}^{-1}(T)$ data is in good agreement
with $\mu_\mathrm{eff} = 0.41\ \mu_\mathrm{B}$/f.u.\ arising from
a $\Gamma_{7}$ ground state and is much less than
$\mu_\mathrm{eff} = 0.77\ \mu_\mathrm{B}$/f.u.\ associated with a
$\Gamma_{8}$ ground state.  However, a linear fit to
$M_\mathrm{S}^{1/\beta}(T)$ (Fig.\ \ref{MsXo}b) resulted in a
value of $0.087(1)\ \mu_\mathrm{B}$/f.u. for $M_\mathrm{sat}$ at
$T = 0$ K, which is $\sim 37\%$ of \mbox{$M_\mathrm{sat} =
g_{J}\langle J_{z} \rangle $} for a $\Gamma_{7}$ ground state at
$T = 0$ K.  In addition, the failure of CEF-based fits to
adequately describe the $\chi_\mathrm{dc}^{-1}(T)$ data may
suggest that the CEF-based fit to $C/T$ is inappropriate. The
Schottky-like anomaly at $10$ K, taken to be a strong indicator of
crystal field splitting, may conceivably be due to a complex
temperature dependence of $\gamma$, which can occur when the value
of $\gamma$ is greatly enhanced,\cite{Stewart84} as is observed in
\SmOsSb{}.

\subsection{Weak Ferromagnetism}
The low value of the magnetic entropy at the transition, the lack
of features in the electrical resistivity at the transition, the
low non-saturating magnetization, and the unexplained critical
exponents seem to suggest that the ferromagnetism is either
unconventional, or due to an unknown impurity phase. Although an
impurity phase may be responsible for the ferromagnetism, it
should be noted that many compounds exhibit similar behavior,
including the itinerant ferromagnets ZrZn$_{2}$, Ni$_{3}$Al, and
Sc$_{3}$In,\cite{Foner67,Ogawa67,deBoer69,Grewe89} many other
heavy fermion systems such as CeNi$_{0.875}$Ga$_{3.125}$, YbRhSb,
and Ce$_{5}$Sn$_{3}$,\cite{Sampathkumaran93,Muro04,Lawrence91} and
the heavy fermion filled skutterudite \SmFeP{}.\cite{Takeda03a}
For instance, the low value of the magnetic entropy at the
transition ($\sim 0.016R \ln 2$) and the small size of the feature
due to magnetic ordering in $C/T$ of \SmOsSb{} are consistent with
the behavior observed in weak itinerant ferromagnets such as
Sc$_{3}$In.\cite{Ikeda81,Ikeda83} Similar behavior is also
observed in several heavy fermion systems such as
CeNi$_{0.875}$Ga$_{3.125}$, YbRhSb, Ce$_{5}$Sn$_{3}$, and
\SmFeP{}.\cite{Sampathkumaran93,Muro04,Lawrence91,Takeda03a} These
heavy fermion systems all show anomalies in $C/T$, although the
magnetic entropy at the phase transitions of each of these systems
ranges from ($0.1$ to $0.35$)$R \ln 2$. The lack of features in
the electrical resistivity at the transition is also consistent
with the weak itinerant ferromagnets ZrZn$_{2}$, Ni$_{3}$Al, and
Sc$_{3}$In,\cite{Ogawa76,Fuller92,Masuda77} as well as
\SmFeP{}.\cite{Takeda03a} None of these compounds shows a clear
indication of a phase transition in $\rho(T)$. In the case of the
magnetization of \SmOsSb{}, the low non-saturating moment is
similar to behavior that has been observed in the magnetization of
\SmFeP{}, which at $1.8$ K reaches a value around $0.14\
\mu_\mathrm{B}/$f.u. by $5$ T.\cite{Takeda03a} This non-saturating
moment is also observed in weak itinerant ferromagnets, such as
ZrZn$_{2}$, Ni$_{3}$Al, and
Sc$_{3}$In.\cite{Foner67,Ogawa67,deBoer69,Grewe89}. The behavior
of the systems described above is not clearly understood and has
been attributed to either Kondo screening or itinerant electron
magnetism.

Assuming the presence of a ferromagnetic impurity phase, the
percentage of this phase present in a sample of \SmOsSb{} can be
estimated from the observed magnetization. Since Sm is the only
magnetic constituent in \SmOsSb{}, it is reasonable to assume a
moment of $\sim 1\ \mu_\mathrm{B}$/f.u. for the impurity phase. In
order to produce the magnetization observed at $2$ K and $5$ T,
the impurity phase would need to constitute $\sim 1\%$ of the
total mass of the crystal, assuming that the molecular weight of
the impurity phase is comparable to that of Sm. Based on this, it
is clear that it would be very difficult to resolve the potential
impurity phase in x-ray powder diffraction data. To check whether
the unconventional magnetic behavior was due to impurities in the
raw materials, single crystals were grown from ultra-high quality
materials (Sm $99.99\%$, Os $99.999\%$, Sb $99.999\%$).
Magnetization measurements made on these crystals did not differ
significantly from those made on previous crystals grown from
lower quality starting materials. This finding seems to indicate
that if there is an impurity phase, then a Sm compound is the most
likely candidate.

The applicability of the modified Arrott analysis to \SmOsSb{}
indicates that the ferromagnetic ordering is not described by mean
field critical exponents ($\beta = 0.5, \delta = 3$, and $\gamma =
1$). Deviation from mean-field behavior is not uncommon. For
instance, the magnetic interactions in a given compound may be
best described by other models, such as the 3D Heisenberg model
for which $\beta = 0.3639, \delta = 4.743$, and $\gamma =
1.3873$.\cite{Chen93} The critical exponents determined for
\SmOsSb{} clearly do not agree with those of the 3D Heisenberg
model, or any of the other standard models.\cite{Binney92} Other
possible explanations for the unconventional critical exponents
include CEF splitting,\cite{Neumann95} disorder of a single
magnetic phase,\cite{Aharoni86,Yeung86} or the existence of
multiple magnetic phases.  In the case of \SmOsSb{}, it seems that
the presence of an impurity phase could lead to the unusual
critical exponents that have been observed.  For example, the
usual mean field critical exponents of some ferromagnetic impurity
could be obscured by the presence of a large paramagnetic
contribution. However, it is also possible that the odd magnetic
behavior is intrinsic to \SmOsSb{} and the unexplained critical
exponents may reflect some type of novel magnetic ordering.
Neutron scattering measurements could offer insight into this
possibility.

\section{Summary}
Measurements of $C(T)$, $\rho(T,H)$, $M(H,T)$, and
$\chi_\mathrm{ac}(T)$ have been performed on the filled
skutterudite compound \SmOsSb{}. The $C(T)$ measurements reveal a
strongly enhanced electronic specific heat coefficient $\gamma
\approx 880$ mJ/mol K$^{2}$, indicative of a large quasiparticle
effective mass $m^{*} \approx 170\ m_\mathrm{e}$.  A fit of a
possible Schottky anomaly to the $C(T)$ data indicates CEF
splitting of the Sm$^{3+} J = 5/2$ six-fold degenerate Hund's rule
multiplet into a $\Gamma_{7}$ doublet ground state and a
$\Gamma_{8}$ quartet excited state separated by $38$ K. The
electrical resistivity has a strong temperature dependence below
$\sim 50$ K. CEF fits to this region agree well with the specific
heat measurements and yield an energy splitting of $\sim 33$ K
between the $\Gamma_{7}$ doublet ground state and the $\Gamma_{8}$
quartet excited state. The resistivity of \SmOsSb{} increases with
field at all temperatures. Below $\sim 10$ K and up to $4$ T, the
resistivity exhibits Fermi-liquid behavior. Fits to the
$\chi_\mathrm{dc}^{-1}(T)$ data using a temperature independent
Van Vleck term yield a value of $\mu_\mathrm{eff} = 0.634\
\mu_\mathrm{B}$/f.u. with an energy gap $\Delta{} = 854$ K between
the $J = 5/2$ and the $J = 7/2$ multiplets.  The hysteresis
observed in $M(H)$ at $2$ K, the low value of the ordered moment,
and conformity of the $M(H,T)$ data to a modified Arrott plot are
consistent with weak ferromagnetic order possibly due to an
impurity phase. Analysis using scaling theory and modified Arrott
plots yield the values $T_\mathrm{C} \approx 2.60(5)$ K and a
spontaneous magnetization at $0$ K of $0.087\ \mu_\mathrm{B}$/f.u.

\section*{Acknowledgements}

We would like to thank S. K. Kim, A. Thrall, and J. R. Jeffries
for experimental assistance. Research at UCSD was supported by the
U. S. Department of Energy under Grant No.~DE-FG02-04ER46105, the
U.S. National Science Foundation under Grant No.~DMR 0335173, and
the NEDO international Joint Research Program. Work at the NHMFL
Pulsed Field Facility (Los Alamos National Laboratory) was
performed under the auspices of the NSF, the State of Florida and
the U.S. Department of Energy.

\newpage
% Table 1
\squeezetable
\begin{table}
\caption{Single crystal structural data for
SmOs$_4$Sb$_{12}$(LaFe$_4$P$_{12}$-type, space group Im$\bar{3}$;
No. 204) taken at $T = 296$\,K, with a scattering angle range of
\mbox{2$^{\circ}$ $< 2 \theta$ $<$ 80$^{\circ}$}.}
\begin{tabular}{|lc|lc|lc|}

 \hline
 \multicolumn{6} {|c|}{\bf SmOs$_4$Sb$_{12}$} \\  \hline
 Crystal size & 84 $\times$ 84 $\times$ 56 $\mu$m$^3$
  & Lattice parameter $a$ [$\rm{\AA}$] & 9.3085(2)
  & Density $\rho$ [g/cm$^3$] & 9.767 \\\hline
 Reflection in refinements & 447 $\leq$ 4 $\sigma$(F$_0$) of 482
  & Number of variables & 11
  & R$_F^2 = \sum \vert F_0^2 - F_c^2 \vert / \sum F_0^2$ & 0.0207 \\\hline
 Goodness of fit &  1.284
  &  &
  &  & \\\hline
 Sm in 2a (0, 0, 0); &
  & Thermal displacements & [$\rm{\AA}^2$]
  & Interatomic distances [$\rm{\AA}$] & \\
 Occupancy & 1.00(2)
  & Sm: $U_{11}$ = U$_{22}$ = U$_{33}$ & 0.0552(6)&
  Sm - 12 Sb & 3.4824 \\\hline
 Os in 8c (1/4, 1/4, 1/4); &
  & Thermal displacements & [$\rm{\AA}^2$]
  & Interatomic distances [$\rm{\AA}$] & \\
 Occupancy & 1.00(1)
  & Os: $U_{11}$ = U$_{22}$ = U$_{33}$ & 0.0019(1) &
  Os - 6 Sb & 2.6241 \\\hline
 Sb in 24g (0, y, z);~~y: & 0.15589(3)
  & Thermal displacements & [$\rm{\AA}^2$]
  & Interatomic distances [$\rm{\AA}$] &  \\
  ~~~~~~~~~~~~~~~~~~~~~~~~~~z: & 0.34009(3)
  & Sb: $U_{11}$ & 0.0019(1)
  & Sb - 1 Sb & 2.9022 \\
  Occupancy & 1.00(1)
  & ~~~~~\,$U_{22}$ & 0.0037(1)
  & ~~~~\,- 1 Sb & 2.9771 \\
 &
  & ~~~~~\,$U_{33}$ & 0.0063(1)
  & ~~~~\,- 2 Os & 2.6241 \\
 &
  &  &
  & ~~~~\,- 1 Sm & 3.4824 \\\hline

\end{tabular}
\label{StructureTBL}
\end{table}
\newpage
\newpage

\end{document}